\documentclass[12pt,preprint]{aastex}

\def\hii{{\rm H}{\scriptsize{\rm II}}}
\def\HII{{\rm H}{\scriptsize{\rm II}}}

\def\um{$\mu$m}

\def\Msun{\mbox{$M_\odot$}}

\def\nodata{$\cdot\cdot\cdot$}

\def\Snu{$S_{\nu}$}

\def\wat{H$_2$O}
\def\amm{NH$_3$}

\def\MCN{\mbox{CH$_3$CN}}
\def\ECN{\mbox{CH$_3$CH$_2$CN}}
\def\hcniso{H$^{13}$CN}

\def\HtCOp{H$^{13}$CO$^+$}

\def\pbeam{beam$^{-1}$}
\def\klambda{\mbox{k$\lambda$}}

\def\Tr{\mbox{$T_{\rm rot}$}}
\def\Te{\mbox{$T_{\rm e}$}}
\def\Tr21{\mbox{$T_{\rm r,21}$}}

\def\Td{\mbox{$T_{\rm d}$}}

\def\Md{\mbox{$M_{\rm d}$}}

\def\lesssim{\mathrel{\hbox{\rlap{\hbox{\lower4pt\hbox{$\sim$}}}\hbox{$<$}}}}
\def\gtrsim{\mathrel{\hbox{\rlap{\hbox{\lower4pt\hbox{$\sim$}}}\hbox{$>$}}}}

\slugcomment{{\small To be appeared in the ApJ\\ Submitted 2004/11/23; Accepted 2005/01/28}}

\shorttitle{Relative Evolutionary Time Scale of Hot Molecular Cores}
\shortauthors{R. S. Furuya et al.}

\begin{document}


\title{Relative Evolutionary Time Scale of Hot Molecular Cores 
with Respect to Ultra Compact HII Regions}


\author{R. S. Furuya\altaffilmark{1}}
\affil{Division of Physics, Mathematics, and Astronomy, California Institute of Technology}
\email{rsf@astro.caltech.edu}

\author{R. Cesaroni\altaffilmark{2}}
\affil{INAF, Osservatorio Astrofisico di Arcetri}
\email{cesa@arcetri.astro.it}

\author{S. Takahashi\altaffilmark{3}}
\affil{Department of Astronomical Science, Graduate University for Advanced Studies}
\email{satoko@nro.nao.ac.jp}

\author{M. Momose\altaffilmark{4}}
\affil{Institute of Astronomy and Planetary Science, Ibaraki University}
\email{momose@mx.ibaraki.ac.jp}

\author{L. Testi\altaffilmark{2}}
\affil{INAF, Osservatorio Astrofisico di Arcetri}
\email{lt@arcetri.astro.it}

\author{H. Shinnaga\altaffilmark{5}}
\affil{Caltech Submillimeter Observatory, California Institute of Technology}
\email{shinnaga@submm.caltech.edu}

\and

\author{C. Codella\altaffilmark{2}}
\affil{INAF -- Istituto di Radioastronomia, Sezione di Firenze}
\email{codella@arcetri.astro.it}


\altaffiltext{1}{1201 East California Boulevard, Pasadena, CA 91125}
\altaffiltext{2}{Largo Enrico Fermi 5, I-50125 Firenze, Italy}
\altaffiltext{3}{Nobeyama Radio Observatory, Nobeyama 411, Minamimaki, Minamisaku, Nagano 384-1305, Japan}
\altaffiltext{4}{Bunkyo 2-1-1, Mito, Ibaraki 310-8512, Japan}
\altaffiltext{5}{111 North A'ohoku Place, Hilo, HI 96720}


\begin{abstract}
Using the Owens Valley and Nobeyama Radio Observatory interferometers,
we carried out an unbiased search for hot molecular cores and 
ultracompact (UC) \hii\ regions toward the high-mass star forming 
region G19.61--0.23.
In addition, we performed 1.2~mm imaging with SIMBA, 
and retrieved 3.5~ and 2~cm images from the VLA archive data base.
The newly obtained 3~mm image brings information on
a cluster of high-mass (proto)stars located in the innermost and densest
part of the parsec scale clump detected in the 1.2~mm continuum.
We identify a total of 10 high-mass young stellar objects:
one hot core (HC) and 9 UC \hii\ regions, 
whose physical parameters are obtained from model fits to
their continuum spectra.
The ratio between the current and expected final radii of the
UC \hii\ regions ranges from 0.3 to 0.9, which leaves
the possibility that all O-B stars formed simultaneously. 
Under the opposite assumption --- namely that star formation occurred randomly ---
we estimate that HC lifetime is less than $\sim$1/3 of that of UC \hii\ regions
on the basis of the source number ratio between them.
\end{abstract}



\keywords{\HII\ regions --- 
ISM: evolution ---
ISM: individual (G19.61--0.23) ---
stars: early type --- 
radio continuum: ISM}


\section{Introduction}

It is widely accepted that high-mass stars ($>8$\Msun) form in clusters.
However, identification of cluster members is complicated because 
zero-age main-sequence stars and high-mass protostars
are deeply embedded in their molecular surroundings.
It is hence impossible to use a simple SED classification analogous 
to that applied to low-mass young stellar objects (YSOs).
In this context, 
ultracompact (UC) ($\lesssim 0.1$ pc) \hii\ regions and 
hot cores (HCs) play a fundamental role
(e.g., Kurtz et al. 2000).  
The former are very young \hii\ regions ionized by an O-B star 
which is embedded in the natal cloud.
The latter are dense, hot, compact molecular cores which host even younger (proto)stars,
although their life time has not been established.
A comparison between the estimated number of UC \HII\ regions in the
Galaxy and the O-B star formation rate yields a duration for the ultracompact
phase of $\sim 10^5$ years (Wood \& Churchwell 1989 [hereafter WC89])
which is two orders of magnitude larger than
expected from the classical expansion model for \HII\ regions
(e.g., Spitzer 1978).
Several models have been proposed to explain this discrepancy
(e.g., Akeson \& Carlstrom 1996, and references therein).
Clearly, one must have better knowledge of properties of clusters to which
HCs and UC \hii\ regions belong and of their natal molecular clumps.\par

A well-studied example of high-mass (proto)star cluster is W49N (at a distance of 11.4~kpc).
Subsequent to the identification of 12 UC \HII\ regions powered by O stars
(De Pree et al. 2000),  
Wilner et al. (2001) identified 6 HCs from an unbiased survey of line and 1.4-mm continuum emission. 
Using the HC to UC\HII\ region number ratio, 
they obtained an estimate of 25--100\% for the lifetime ratio between the
HC and UC \hii\ region phases.
However, the W49N study has an important limitation that
the region is very distant which sets limits on the census of detectable YSOs.
The high-mass star forming region G19.61--0.23 is an excellent target for a study of cluster formation
because of its distance ($d=3.5$ kpc; Churchwell, Walmsley, \& Cesaroni 1990), and
potential richness in terms of YSOs as indicated by studies at cm 
(e.g., 
Garay, Reid, \& Moran 1985 [hereafter GRM85]; 
WC89; 
Garay et al. 1998; 
Forster \& Caswell 2000) 
and mid-infrared (MIR) (De Buizer et al. 2003) wavelengths.
In this letter, 
we present a study of the cm continuum emission with higher sensitivity and better angular resolution, 
complemented by mm wavelength observations.

\section{Observations and Archive Data Retrieval}

Aperture synthesis observations at 3~mm were carried out using Owens
Valley Radio Observatory (OVRO) millimeter array
and Nobeyama\footnote{Nobeyama Radio Observatory is a branch of
the National Astronomical Observatory, operated by the Ministry of
Education, Culture, Sports, Science and Technology, Japan.} Millimeter Array (NMA)
with a total of 6 array configurations in the period from 2002 December till 2004 April.
More than 15 molecular lines (R. Furuya et al. in preparation) were observed
simultaneously with the continuum emission at 89.9~GHz.
The flux density of the calibrator was bootstrapped from planets with
an uncertainty of 15\%. 
Combining the continuum data from both arrays, we obtain
a synthesized beam of 1\farcs52 $\times$ 1\farcs51, 
and an RMS noise level of 2.8~mJy~beam$^{-1}$.\par

On 2003 July 29, we used the SIMBA bolometer array on the
SEST 15-m telescope to obtain a parsec-scale map 
($3.0\arcmin\times 4.0\arcmin$ in size) of  1.2~mm continuum emission.
Observing $\eta$ Carinae and Uranus, 
we checked pointing accuracy (RMS $\simeq 2\arcsec$),
and made flux calibrations.\par

We retrieved 3.5~ and 2~cm data from the archive of the 
NRAO\footnote{The National Radio Astronomy Observatory
is a facility of the National Science Foundation operated under cooperative agreement
by Associated Universities, Inc.} Very Large Array (VLA).
The 3.5~cm data were taken from projects AW374A and AK470 (A-array),
AK470 (B), AK423 (CnB), and AK450 (D), and the
2~cm data from AW158 (B and C), AK423 (CnB).
Resulting beam sizes and noise levels are
0\farcs33$\times$0\farcs23 and 0\farcs55$\times$0\farcs50
with 0.31 and 1.7 mJy \pbeam\ at 3.5 and 2~cm, respectively.

\section{Results and Analysis}

\begin{figure}
\begin{center}
\includegraphics[width=.42\linewidth,angle=0]{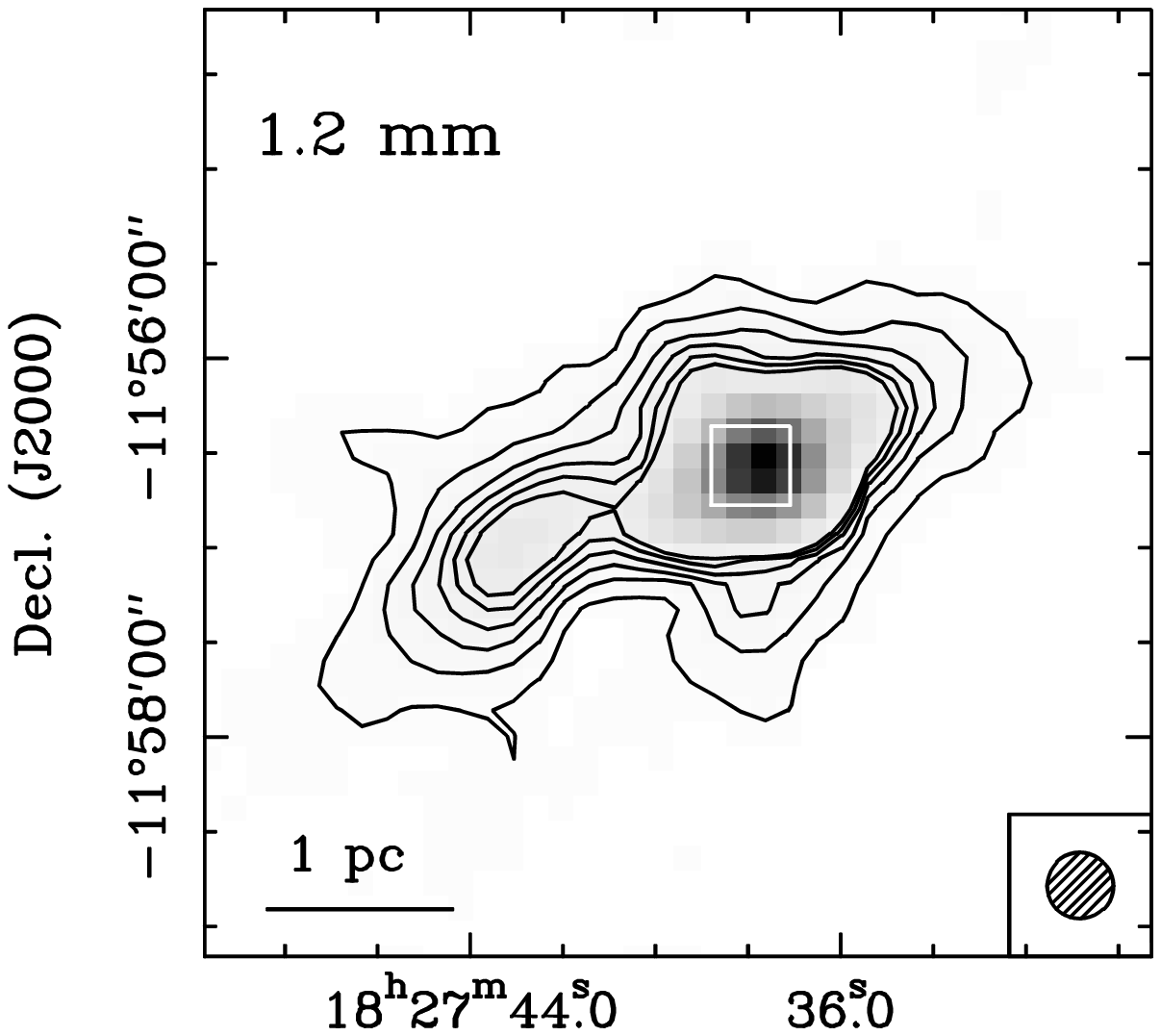}
\end{center}
\caption{Parsec scale view of the cloud hosting the G19.61--0.23 region
in 1.2~mm continuum emission.
Contours start from the 3$\sigma$ level of noise level (30.7 mJy \pbeam\ ), 
and end at the 15$\sigma$ level with 3$\sigma$ steps.
The hatched ellipse in the bottom right corner indicates the SIMBA beam size ($22\arcsec$),
and the central white box marks the region shown in Figure \ref{fig:3mmX}.
}
\label{fig:simba}
\end{figure}

Figure \ref{fig:simba} presents the SIMBA 1.2~mm continuum emission map where
two clumps
separated by $\sim$75\arcsec\ (i.e. 1.3 pc) are seen over a region of
$\sim 4.1$ pc$\times$1.7~pc.
The brighter clump to the west hosts the
G19.61--0.23 complex.  A total flux density (\Snu) of 19.4 Jy is obtained
by integrating the emission inside the 3$\sigma$ contour level.
Figure \ref{fig:3mmX}a shows the VLA image of the 3.5~cm continuum emission
toward the center of the western clump,
where a cometary shaped and several compact \HII\ regions,
surrounded by a 0.1-pc halo, are seen.\par

Since the minimum projected baseline lengths 
(0.78, 7.7, and 2.8 \klambda\ for the 3.5, 2~cm, and 3~mm data, respectively)
and beam sizes ($\S 2$) are not identical,
we reconstructed all the interferometric images with a 1\farcs5 beam 
using visibilities with a {\it uv}-radius greater than 7.7 k$\lambda$.
This makes our images insensitive to structures
more extended than 27\arcsec\ (i.e. 0.46 pc).
Thus only the 0.1-pc halo common to all of the cluster members is resolved out.
Figure \ref{fig:3mmX}b represents an overlay of such maps at 3.5, 2~cm and 3~mm.
Most of the 3~mm peaks are associated with previously identified cm sources
(GRM85; WC89; Garay et al. 1998), suggesting that our 3~mm map is sensitive 
to both free-free emission from ionized gas inside the compact \HII\ regions 
and thermal emission from dust grains in the surrounding cores.
To establish the properties of the dust continuum emission, 
we need to remove the free-free emission from the 3~mm map. 
We produced a ``pure'' free-free emission image at 3~mm by
extrapolating the 3.5~cm map assuming optically thin emission
($I_{\nu}\propto \nu^{-0.1}$).
Then we subtracted this from the 3~mm continuum image, thus
obtaining a ``pure'' dust emission map (see Figure \ref{fig:3mmX}c).
Taking into account the calibration errors at the various wavelengths,
we estimate a subtraction uncertainty of 4.3 mJy \pbeam.
The resulting map
presents a strong ($29\sigma$) peak lying between the UC \HII\ regions labeled A and C (see the yellow cross in Figure \ref{fig:3mmX}a), 
where no emission peak is seen in 2 and 3.5~cm images.
Instead, the 3~mm peak coincides with a molecular core (Figure \ref{fig:3mmX}d) detected 
in the \ECN\ (11--10) and 
SO$_2$ (8$_{3,5}-9_{2,8}$) lines: these molecular species
are typical tracers of HCs (van Dishoeck \& Blake 1998).
The core is also traced by other molecules 
such as \amm\ (Garay et al. 1998), 
\MCN\ (Kurtz et al. 2000),
HCOOCH$_3$,
\hcniso, 
HN$^{13}$C, 
\HtCOp, and SO (R. Furuya et al. in preparation).
Such a richness in molecular species makes this core
very similar to Orion-KL (Wright, Plambeck, \& Wilner 1996) and other HCs
(Kurtz et al. 2000).
In addition, 
the presence of \wat\ (Hofner \& Churchwell 1996) and OH (GRM85) masers
strongly suggest that the (proto)star deeply embedded in the core is too young to ionize 
its surrounding material 
because masers are believed to disappear 
during the development of an UC \HII\ region.
No other prominent ``dust'' condensation was detected
above a 5$\sigma$ upper limit of 21.5~mJy \pbeam, although extended emission
is seen towards some of the \HII\ regions. Such an upper limit
corresponds to a sensitivity in gas plus dust mass of 27 \Msun\ \pbeam\
for a typical HC temperature of 100~K (Kurtz et al. 2000).
The large majority ($\sim$85\%) of the HCs known to date 
(e.g. Table~1 of Kurtz et al. 2000) are above this limit, 
but we cannot rule out the possibility to
have missed the least massive ones ($\sim$10~$M_\odot$).\par

We have identified the UC \hii\ regions in the cluster with
a procedure, which iteratively fitted and subtracted prominent
emission peaks from the cm maps, until the signal in the residual image
was below 5$\sigma$.
The borders of the cometary shaped UC \hii\ region A and the source J at 2 cm were 
identified as the 5$\sigma$ contour levels enclosing them, 
whereas for all the other UC \hii\ regions a 2-D gaussian fit was used.
This gave us the
peak position, \Snu, and deconvolved source diameter of the sources.
In the estimate of the error on \Snu, we considered
the formal error of the fit, the flux calibration uncertainty, and
the contribution due to the noise in the map.
In this way, we identified 9 UC \hii\ regions, labeled A--D, F, G, I, J, and K. 
All of these were identified both at 3.5 and 2~cm, with the sole exception
of J, too faint to be detected in the 2~cm image.

In order to derive the properties of the UC \HII\ regions at 3~mm, it was
necessary to disentangle the HC from UC \HII\ regions A and C. 
For this purpose, we used the free-free subtracted map at 3~mm (Figure \ref{fig:3mmX}c)
to fit the HC with a 2-D Gaussian.
Then this gaussian model was subtracted
from the original 3~mm image. 
The resulting ``clean'' image was treated
in the same way as the 3.5 and 2~cm maps to identify the counterparts of
the cm sources.
Because of the difficulty to define unique area for object F at 3~mm,
we estimate an uncertainty equal to the difference between the flux estimate
obtained from Gaussian fitting and that from
integrating the emission inside the 5$\sigma$ contour.
In Figures \ref{fig:3mmX}b and c, one can see that an isolated 3~mm source lies
to the east of A: we assume that this emission is associated
with the tail of the cometary UC \HII\ region A rather than to a
distinct YSO.
We thus identified 6 objects at 3~mm, including the HC, above a detection threshold of 5$\sigma$.
In conclusion, we have identified a total of 10 high-mass YSOs:~
1 HC and 9 UC \HII\ regions (Table \ref{tbl:source}).
Note that we detected all the radio sources previously identified by other
authors, except source E of WC89.
All 9 UC \HII\ regions have been detected at more than 2 bands at
cm wavelengths, if one takes into account identifications reported in
the literature.

\section{Discussion}

We analyzed the continuum spectra of the whole region
and of each cluster member by modeling
the emission as originating from a dust-free \HII\ region surrounded by
a dusty molecular shell. 
Spherical symmetry and homogeneity are assumed for both the \HII\ region and the shell. 
The spectrum of the whole region is shown in Figure~106 of WC89. To the data
reported in this figure we have added our fluxes at 3~ and 1.2~mm ($\S 3$),
and that at 350 $\mu$m by Mueller et al. (2002).
Most of the 3~mm emission is due to free-free, while 
the emission shortward of 1.2~mm is originating from dust. 
Given the complexity of the region seen in free-free emission,
we estimate only a pc-scale mean dust temperature (\Td) and a total mass of the gas plus dust (\Md).
Therefore, the contribution of the \HII\ region was
considered only to subtract the free-free component from the spectrum.
We adopted a dust absorption coefficient
$\kappa_{\nu}=\kappa_0\left(\nu/\nu_0\right)^\beta$ with
$\kappa_0$=0.005~cm$^2$~g$^{-1}$ at $\nu_0$=230~GHz
(Preibish et al. 1993) 
and $\beta$=1.5.
We obtain \Td$\simeq$42~K, 
and \Md$\simeq$2800$\pm$100~$M_{\sun}$, which
are very similar to the values of other clumps containing high-mass YSOs
(e.g. Fontani et al. 2002, and references therein).\par

Subsequently we have analyzed the continuum spectra of the 10 sources in the 
cluster using the above model.
In the fits we fixed radii of \HII\ regions ($R_{\rm HII}$) 
and dusty shells ($R_{\rm d}$) to the values derived respectively from the 3.5~cm and 3~mm maps,
and the electron temperatures (\Te) to those obtained from recombination line observations 
(Garay et al. 1998).
For source A at the 3 bands, 
D and F at 3~mm, 
and J at 2~cm, 
we defined an effective radius as that of a circle with the same area as
the UC \hii\ region or the dusty shell.
We assumed a mean \Te\ of 7200 K for the objects where \Te\ are not available
in Table 5 of Garay et al. 1998.
Therefore, free parameters were the emission measure
($EM$) of the \HII\ region, and \Md\ and \Td\ of the dusty molecular shell.
Our spectral analysis shows that fluxes at 3~mm (Table~\ref{tbl:fit}), 
20.8, 18.1, and 11.7 \um\ (De Buizer et al. 2003) towards objects B, F, I, and the HC
are fitted with combinations of thermal dust plus free-free emission,
implying that they are likely to be embedded in dust cocoons.
Only the HC is not detected at cm wavelengths, 
which indicates that the putative embedded stars (or star) are too
young or not massive enough to develop an UC \HII\ region.
We favor the idea that the HC contains very young massive (proto)star(s) as it
is in all aspects (richness of the molecular composition, 
large mass of $800~M_{\sun}$, 
high temperature of 65~K) 
similar to the other HCs known in the literature, 
which are believed to be the cradle of early-type stars (Kurtz et al. 2000).
If the HC contains a deeply embedded O-B star,
one may estimate the maximum radius of the associated \HII\
region from the sensitivity of the 3.5~cm image.
Assuming that the free-free emission is optically thick, 
such a radius must be smaller than 2.4$\times 10^{-3}$ pc.
The continuum spectrum of
D is rising from cm wavelengths to 3~mm 
(Table~\ref{tbl:fit}), but does not show MIR emission.
This sets upper limits on \Td\ and \Md.
The 3~mm flux of source A (589$\pm$115~mJy) is only marginally above the
value expected (409~mJy) from extrapolation of the free-free cm fluxes.
Considering the uncertainties and the fact that 
its effective $R_{\rm d}$ is smaller than that of $R_{\rm HII}$, 
we believe that thermal dust emission is not dominant at 3~mm.
Finally, the spectrum of sources C, G, J, and K can be fitted with free-free emission only, 
suggesting that they have already dispersed most of the surrounding
material.\par

An unbiased survey such as ours could be used to estimate the
lifetime of HCs with respect to that of UC \HII\ regions on the basis of 
number ratio between the two. 
However, this applies 
only to a series of star formation episodes randomly distributed in time,
and not if one is observing the result of a coherent burst of star formation 
(e.g. Welch et al. 1987).
While in the former case the different numbers of sources in different
evolutionary stages would be directly related to the time spent in each phase,
in the latter hypothesis such numbers would be determined by the local
physical conditions around individual stars.
It is hence important to find out 
whether the ages of our sources are spread over a sufficiently large time interval 
or all very similar.\par

Naively, one might expect that the age of an \HII\ region is strictly
related to its diameter. 
However the correspondence between size and age is still matter of debate.
Nevertheless, in most expansion models of an \HII\ region is related 
only to the spectral type of the star and to the properties of the surrounding environment. 
It is hence reasonable to assume that the expansion will come to
an end when pressure equilibrium is reached between the ionized and the surrounding molecular gas. 
As illustrated by De Pree, Goss, \& Gaume (1998), 
their eqs.~(4) and (7) give the final \HII\ region radius 
($R_{\rm HII}^{\rm final}$) 
as a function of the stellar Lyman continuum photons,
\Te\ ,
and the density and temperature of the molecular gas.  
Here we derived spectral types of 
B0 for A, B0.5 for B-D, F, G, I, K, and B1 for J.
The value of $R_{\rm HII}^{\rm final}$ should be compared to the current UC \HII\ radius measured from the maps.  
Although it is impossible to obtain an ``absolute'' estimate of the age,
ratios of $R_{\rm HII}/R_{\rm HII}^{\rm final}$ 
may be used to estimate ``relative'' ages of UC \HII\ regions, 
with lower ratios corresponding to younger objects.
To compute $R_{\rm HII}^{\rm final}$, 
we used the corresponding values of $\Te$ and $\Td$, and a mean density derived from
the SIMBA observations ($\sim 3\times 10^6$ cm$^{-3}$).
The ratio ranges from 0.3 to 0.9 (see Table~\ref{tbl:fit}). 
If the UC \hii\ regions have not yet reached pressure equilibrium with the surrounding cloud, 
such a ratio seems to indicate that they are in similar evolutionary stages,
thus supporting the ``starburst'' scenario.\par

However, several caveats are in order: 
the density of molecular gas may vary over the natal cloud and the Lyman
continuum of the stars may have been underestimated because of optically
thick free-free emission and dust absorption inside the \HII\ region.
This may easily introduce an uncertainty of an order of magnitude on
$R_{\rm HII}^{\rm final}$ thus making the ``starburst'' hypothesis
questionable. 
In fact, the expansion timescales obtained from eq.~(3) of 
De Pree, Goss, \& Gaume (1998) span a small range (120--480~yr), 
suggesting that the UC \hii\ regions are close to pressure equilibrium.
It is hence worth considering also the opposite scenario,
and take the number ratio between HCs and UC \HII\ regions as an estimate of
their relative lifetimes.\par

Considering Poisson statistical errors, the number of HCs and UC \hii\ regions
are respectively 1$\pm$1 and 9$\pm$$3$, so that their ratio is
$11^{+22}_{-11}$\%.
This implies that the HC phase ($\tau_{\rm HC}$) should last less than
$\sim$1/3 of the UC \HII\ region phase ($\tau_{\rm H\scriptsize II}$), 
with a probability of 4\% that the ratio exceeds this value.
This conclusion seems to differ from that inferred for the O-star cluster in
W49N (Wilner et al. 2001), where
$\tau_{\rm HC}/\tau_{\rm H\scriptsize II}\simeq25$--100\%.
The two results could indicate that
$\tau_{\rm H\scriptsize II}\simeq 3$--$4\, \tau_{\rm HC}$.
Alternatively, the difference might be easily explained in the ``starburst''
scenario,
where the ratio between HCs and UC \HII\ regions depends on
the environment around each star. 
Finally, another caveat is in order. Our statistics might be affected by the limited
sensitivity of our 3~mm map. As previously explained, we are
sensitive to HCs with masses in excess of $\sim$27~$M_\odot$, whereas
at cm wavelengths we detect stars earlier than B1. If O stars are born in
massive HCs and B stars in lower mass HCs, our comparison between
UC \hii\ regions around B stars and HCs hiding O stars is bound to
be unreliable.
More numerous and sensitive interferometric observations of O-B star 
clusters will be needed to improve the statistics.


\acknowledgments

The authors gratefully acknowledge an anonymous referee whose comments 
significantly improved the quality of the paper.
R. S. F. thanks Prof. A. I. Sargent and J. M. Carpenter for critical readings of the manuscript
and encouragement, M. Nielbock and L. Haikala for their generous help at
SIMBA observations and data reduction.  
Many thanks are due also to T. Saito for his early contribution to this study, 
and to the staff at OVRO, NRO, NRAO, and SEST.
Research at the Owens Valley Radio Observatory 
is supported by the National Science Foundation through NSF grant AST 02-28955.

\begin{figure}
\caption{(a) 3.5~cm continuum emission map taken with VLA. 
The hatched ellipse in the bottom right corner indicates the beam size.
The white contour corresponds to the 10$\sigma$ level.
The yellow cross indicates the position of the {\it dust} continuum emission peak 
(R.A.$=18^{\rm h}27^{\rm m}38.^{\rm s}066$,
Decl.$=-11\degr ~56\arcmin ~37\farcs20$ in J2000).
(b) Overlay of 3~mm continuum emission map taken with OVRO and NMA 
(grey scale plus black contour) on the 3.5~cm ({\bf thick} purple contours) and 2~cm 
(thin green contours) maps CLEANed with the 3~mm beam (see text).
The labels denote identified sources.
Contour levels for the 3~mm map are 5$\sigma$ intervals starting from the
5$\sigma$ level ($\sigma_{\rm 3mm} =4.3$ mJy \pbeam\ ). 
For the sake of clarity,
we plot only 5$\sigma$, 10$\sigma$, and
30$\sigma$ levels for the 3.5~cm map, and 5$\sigma$, 10$\sigma$,
20$\sigma$, and 40$\sigma$ levels for the 2~cm map,
where $\sigma =$0.38 and 1.6 mJy \pbeam\ , respectively.
(c) ``Pure'' {\it dust} continuum emission map obtained after subtraction
from the 3~mm image of the 3.5~cm map suitably scaled to 3~mm (see text).
Contours start from the 5$\sigma$ level with a 5$\sigma$ step ($\sigma_{\it dust} =4.3$ mJy \pbeam\ ).
(d) Overlay of total integrated intensity maps of  \ECN\ (11--10) (grey scale) and
SO$_2$ $8_{3,5}-9_{2,8}$ emission (green contours) taken with the NMA.
Both the grey scale and green contours start from the 5$\sigma$ level with
5$\sigma$ steps, where noise levels and beam sizes are
14.1 mJy \pbeam\ and 6\farcs40$\times$3\farcs65 for \ECN, and
9.4 mJy \pbeam\ and 4\farcs59$\times$2\farcs95 for SO$_2$.
The open red circles and 6 blue triangles indicate the positions of the 
OH (GRM85), and 
\wat\ (Hofner and Churchwell 1996)
maser spots, respectively.
The yellow cross has the same meaning as in (a).
}
\label{fig:3mmX}
\end{figure}

\begin{figure}
\begin{center}
\includegraphics[width=.5\linewidth,angle=0]{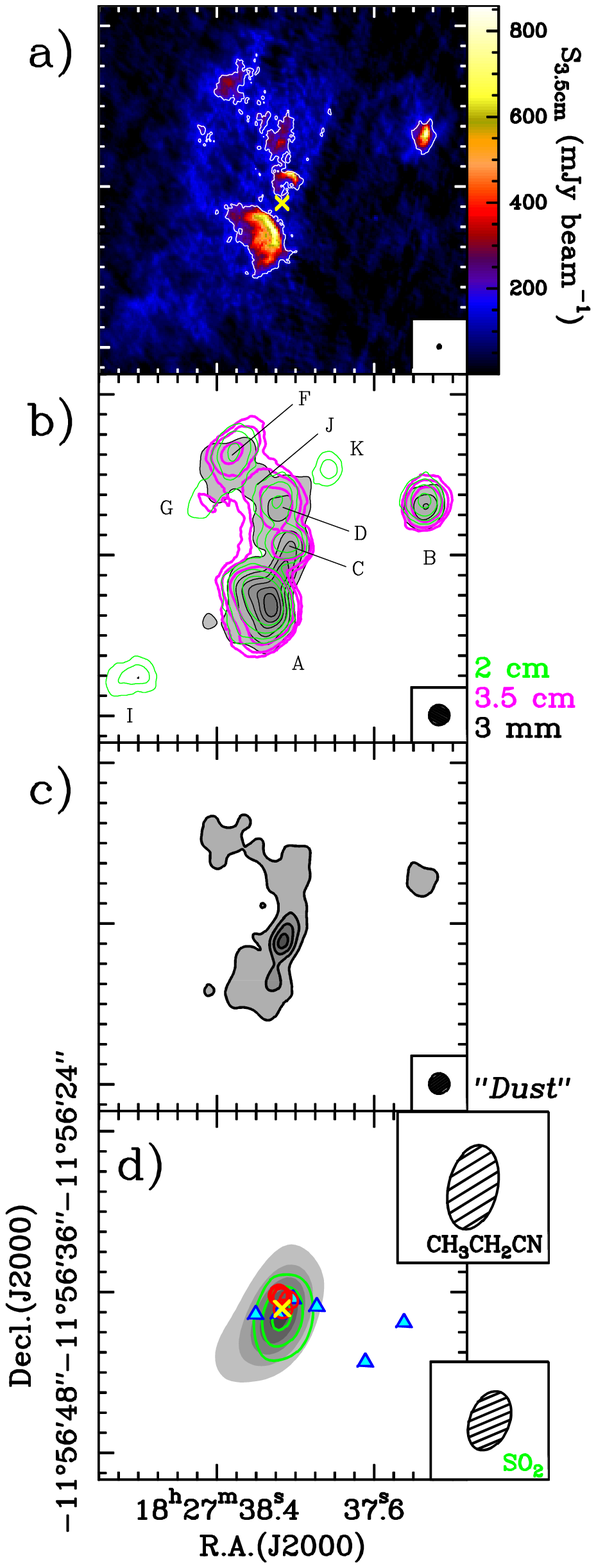}
\end{center}
\end{figure}

\clearpage

\begin{deluxetable}{ccllclllrccrrc}
\setlength{\tabcolsep}{0.02in}
\tabletypesize{\footnotesize}
\tablecaption{Parameters of Identified Sources\label{tbl:source}}
\tablewidth{0pt}
\tablehead{
\multicolumn{4}{c}{YSO Name\tablenotemark{a}} & &
\colhead{$S_{8.43}\tablenotemark{b}$} &
\colhead{$S_{14.9}\tablenotemark{b}$} &
\colhead{$S_{89.9}\tablenotemark{b}$} &
\colhead{$R_{\rm HII}\tablenotemark{c}$} & 
\colhead{$EM$} & 
\colhead{$R_{\rm d}$} &
\colhead{$T_{\rm d}$} & 
\colhead{$M_{\rm d}$} & 
\colhead{$R_{\rm HII}/R_{\rm HII}^{\rm final}$} \\
\cline{1-4}
\cline{6-8}
\colhead{} & 
\colhead{WC89} &
\colhead{G98} &
\colhead{D03} & &
\colhead{(mJy)} &
\colhead{(mJy)} &
\colhead{(mJy)} &
\colhead{(pc)} & 
\colhead{(pc cm$^{-6}$)} & 
\colhead{(pc)} &
\colhead{(K)} & 
\colhead{($M_{\sun}$)} &
\colhead{} 
}
\startdata
A  & A        & A  & 3 & & 442$\pm$88        & 521$\pm$104          & 589$\pm 115$           & 5.2E--2\tablenotemark{d}    & 1.3E+7  & 4.5E--2\tablenotemark{d}  & 70                      & 760                  & 0.8 \\
B  & B        & B  & 2 & & 72$^{+16}_{-21}$  & 62$^{+15}_{-12}$     & 201$^{+40}_{-73}$      & 7.8E--3                     & 8.3E+7  & 2.0E--2                    & 70                     & 400                     & 0.3 \\
C  & C        & F  &   & & 52$\pm$10         & 50$^{+38}_{-10}$     & 56$^{+26}_{-11}$       & 1.2E--2                     & 2.5E+7  & \nodata                    & \nodata                & \nodata                 & 0.4 \\
D  & D        &    &   & & 68$^{+14}_{-20}$  & 49$^{+40}_{-10}$     & 187$^{+37}_{-48}$      & 2.0E--2                     & 1.2E+7  & 3.2E--2\tablenotemark{d}   & $<70$\tablenotemark{e} & $<580$\tablenotemark{e} & 0.6 \\
F  &          & CD & 4 & & 68$^{+72}_{-33}$  & $57^{+29}_{-11}$     & 153$^{+31}_{-46}$      & 1.7E--2                     & 1.5E+7  & 2.7E--2\tablenotemark{d}   & 70                     & 480                     & 0.6 \\
G  &          &    &   & & 5.0$\pm$1         & $24^{+6}_{-7}$       & $<17$\tablenotemark{e} & 2.0E--2                     & 4.4E+6  & \nodata                    & \nodata                & \nodata                 & 0.8 \\
HC &          &    &   & & $<1.8$            & $<6.0$               & 147$\pm$29             & $<2.4$E--3\tablenotemark{f} & \nodata & 1.7E--2                    & 65                     & 800                     & \nodata \\
I  &          & E  & 5 & & (4.7$^{+1}_{-3}$) & $43^{+22}_{-9}$      & (21$\pm$4)             & 2.2E--2                     & 6.3E+6  & $\sim$3E--2\tablenotemark{g} & (70)                 & (35)                    & -- \\
J  & & CD\tablenotemark{h}   &   & & 5.2$\pm$1  & $<6.0$            & $<17$                  & 1.2E--2\tablenotemark{d}    & 2.5E+6  & \nodata                    & \nodata                & \nodata                 & 0.9 \\
K  & D\tablenotemark{h} & & & & $<1.8$       & $17^{+7}_{-3}$       & $<17$                  & 2.2E--2                     & 2.5E+6  & \nodata                    & \nodata                & \nodata                 & -- \\
\enddata
\tablenotetext{a}{From the left, source labels identified in this work, WC89, Garay et al.(1998), and De Buizer et al. (2003).}
\tablenotetext{b}{Flux densities are measured with a 1\farcs5 beam and a minimum projected baseline length of 7.7 k$\lambda$ ($\S 3$), 
and those in parenthesis are marginal detections that the significance levels of the peak intensities range 
between $\ge 3\sigma$ and $< 5\sigma$.
All the results in the following analysis based on the marginal detections are shown in parenthesis.
Upper limits are calculated with the 10$\sigma$ level noise levels over the 1\farcs5 beam.}
\tablenotetext{c}{Measured at 14.9 GHz for the sources I, G, and K.}
\tablenotetext{d}{Effective radius ($\S 4$)}
\tablenotetext{e}{Upper limit is from lack of well-defined emission peak at the 18.1 \um\ map (Figure 8 of De Buizer et al. 2003)}
\tablenotetext{f}{Assuming optically thick free-free emission ($\S 4$)}
\tablenotetext{g}{Obtained from the 18.1 \um\ map (Figure 8 of De Buizer et al. 2003)}
\tablenotetext{h}{The source J is located inside a circle given by the position of the source CD in G98 and its angular size 
(see their Table 2). The source K corresponds to the north-western peak of the 6~cm source D of WC89.}
\label{tbl:fit}
\end{deluxetable}




\end{document}